\pdfoutput=1
\documentclass[aps,prd,reprint,twocolumn,showpacs,superscriptaddress,nofootinbib]{revtex4-1}  
\usepackage{tabularx}
\makeatletter
\def\hlinewd#1{%
\noalign{\ifnum0=`}\fi\hrule \@height #1 %
\futurelet\reserved@a\@xhline}
\makeatother
\usepackage{scrextend}
\usepackage{amsmath}
\usepackage{amssymb}
\usepackage{epsfig}
\usepackage{graphicx}
\usepackage{hyperref}
\usepackage[normalem]{ulem}
\usepackage{dcolumn}   
\usepackage{slashed}
\usepackage{color}
\usepackage{rotating}
\usepackage[margin=0.9in,a4paper]{geometry}
\usepackage[table,xcdraw,dvipsnames]{xcolor}
\usepackage[utf8]{inputenc}
\usepackage{colortbl}
\usepackage[normalem]{ulem}
\usepackage{mathrsfs} 
\usepackage[version=4]{mhchem}
\usepackage{cancel}

\definecolor{nicered}{rgb}{0.7,0.1,0.1}
\definecolor{nicegreen}{rgb}{0.1,0.5,0.1}
\definecolor{red}{rgb}{1.0, 0, 0}
\definecolor{CYAN}{rgb}{0, 1, 1} 

\newcommand{\X}{{\cal X}}

\renewcommand{\[}{\left[}
\renewcommand{\]}{\right]}
\renewcommand{\(}{\left(}
\renewcommand{\)}{\right)}

\newcommand{\bdm}{\begin{displaymath}}
\newcommand{\edm}{\end{displaymath}}
\newcommand{\bea}{\begin{eqnarray}}
\newcommand{\eea}{\end{eqnarray}}

\renewcommand{\X}{\mathcal{X}}

\definecolor{nicered}{rgb}{0.7,0.1,0.1}
\definecolor{nicegreen}{rgb}{0.1,0.5,0.1}
\definecolor{red}{rgb}{1.0, 0, 0}
\definecolor{niceblue}{rgb}{0,0,0.8}
\definecolor{red}{rgb}{1.0, 0, 0}
\hypersetup{colorlinks,citecolor= nicegreen,linkcolor= nicered,urlcolor=nicered}

\def\eq#1{{Eq.~(\ref{#1})}}
\def\eqs#1#2{{Eqs.~(\ref{#1})--(\ref{#2})}}

\def\sect#1{{Sect.~\ref{#1}}}

\def\vev#1{\left\langle #1\right\rangle}
\def\abs#1{\left| #1\right|}

\def\gsim{\raise0.3ex\hbox{$\;>$\kern-0.75em\raise-1.1ex\hbox{$\sim\;$}}}
\def\lsim{\raise0.3ex\hbox{$\;<$\kern-0.75em\raise-1.1ex\hbox{$\sim\;$}}}

\def\mb[#1]{\mathbf{#1}}

\renewcommand{\bar}{\overline}

\definecolor{LightCyan}{rgb}{0.88,1,1}
\definecolor{piggypink}{rgb}{0.99, 0.87, 0.9}
\definecolor{applegreen}{rgb}{0.55, 0.71, 0.0}
\definecolor{darkpastelgreen}{rgb}{0.01, 0.75, 0.24}
\definecolor{green-yellow}{rgb}{0.68, 1.0, 0.18}

\newcommand{\beq}{\begin{equation}}
\newcommand{\eeq}{\end{equation}}
\newcommand{\beqa}{\begin{eqnarray}}
\newcommand{\eeqa}{\end{eqnarray}}

\newcommand{\GeV}{{\, \rm GeV}}

\begin{document}



\title{Do Finite Density Effects 
Jeopardize Axion Nucleophobia in Supernovae? 
}

\author{Luca Di Luzio}
\email{luca.diluzio@pd.infn.it}
\affiliation{\small \it Istituto Nazionale di Fisica Nucleare, Sezione di Padova, Via F.~Marzolo 8, 35131 Padova, Italy}
\author{Vincenzo Fiorentino} 
\email{vincenzo.fiorentino@uniroma3.it}
\affiliation{\small \it Dipartimento di Fisica e Astronomia `G.~Galilei', Universit\`a di Padova, Via F.~Marzolo 8, 35131 Padova, Italy}
\affiliation{\small \it Istituto Nazionale di Fisica Nucleare, Laboratori Nazionali di Frascati, 00044 Frascati, Italy}
\affiliation{\small \it Universit\`a degli Studi Roma Tre, Via della Vasca Navale 84, I-00146, Rome, Italy}
\author{\\ Maurizio Giannotti} 
\email{mgiannotti@unizar.es}
\affiliation{\small \it Centro de Astropartículas y Física de Altas Energías, University of Zaragoza, Zaragoza, 50009, Aragón, Spain}
\affiliation{\small \it Department of Chemistry and Physics, Barry University, Miami Shores, 33161, FL, United States of America}
\author{Federico Mescia}
\email{federico.mescia@lnf.infn.it}
\affiliation{\small \it Istituto Nazionale di Fisica Nucleare, Laboratori Nazionali di Frascati, 00044 Frascati, Italy} 
\affiliation{\small \it On leave of absence from Universitat de Barcelona}
\author{Enrico Nardi} 
\email{enrico.nardi@lnf.infn.it}
\affiliation{\small \it Istituto Nazionale di Fisica Nucleare, Laboratori Nazionali di Frascati, 00044 Frascati, Italy}
\affiliation{\small \it Laboratory of High Energy and Computational Physics, HEPC-NICPB,  R\"avala 10, 10143 Tallinn, Estonia}

\begin{abstract}
\noindent
Nucleophobic axion models, wherein axion couplings to both protons and neutrons 
are simultaneously suppressed, can relax the stringent constraints from SN 1987A. However, it remains uncertain whether these models maintain their nucleophobic property 
under the influence of finite baryon density effects. These are especially relevant in astrophysical environments near saturation density, such as Supernovae (SNe). In this study, we demonstrate that the nucleophobic solution remains viable also at finite density. Furthermore, we show that the SN axion bound relaxes significantly in nucleophobic models, even when accounting for the integration over the non-homogeneous environment of the SN core.

\end{abstract}

 \maketitle



\section{Introduction}
\label{sec:intro}

The Quantum Chromodynamics (QCD) axion 
offers a compelling solution to the strong CP problem \cite{Peccei:1977hh,Peccei:1977ur,Weinberg:1977ma,Wilczek:1977pj} 
and serves as an excellent dark matter candidate \cite{Dine:1982ah,Abbott:1982af,
Preskill:1982cy}, 
thereby addressing two major challenges in 
contemporary 
high-energy physics. 
The axion's model-independent properties are primarily determined by a single parameter, the axion decay constant, $f_a$,  
which determines the value of the axion mass, $m_a$, and the overall suppression of the 
axion couplings to Standard Model (SM) particles. 
The actual strength of a specific coupling, however, crucially depends also on the particular  
axion model under consideration,  
through the Wilson coefficients 
of the axion effective Lagrangian (see e.g.~\cite{DiLuzio:2020wdo}). 
This fact is especially important when addressing astrophysical constraints 
on the axion parameter space, which are often presented 
in terms of benchmark KSVZ \cite{Kim:1979if,Shifman:1979if} and DFSZ \cite{Zhitnitsky:1980tq,Dine:1981rt} axion models. 

In Ref.~\cite{DiLuzio:2017ogq},  
a new class of axion models, termed ``astrophobic'', was shown to simultaneously suppress the axion's couplings to nucleons and electrons, thereby loosening the stringent bounds imposed by 
Supernova (SN) 1987A \cite{Carenza:2019pxu,Carenza:2020cis}, 
as well as by the observed evolution of red giants \cite{Capozzi:2020cbu,Straniero:2020iyi}
and white dwarfs \cite{MillerBertolami:2014rka,Corsico:2019nmr}.
In this case, the astrophysical constraints on the axion mass 
can be relaxed by more than one order of magnitude,  
compared to, for example, the benchmark KSVZ model. 

It should be noted that achieving the 
nucleophobic condition in a QCD axion context is non-trivial, due to the
irreducible contribution to the axion-nucleon coupling arising from the anomalous axion-gluon interaction.
In fact, it can be seen that in benchmark axion models 
it is not possible to simultaneously suppress the axion coupling to both protons and neutrons. 
Remarkably, the nucleophobic conditions necessarily require a non-universal Peccei-Quinn (PQ) charge assignment \cite{DiLuzio:2017ogq}, 
where SM quarks from different generations carry distinct PQ charges, thus establishing an intriguing link with flavour physics. 

The nucleophobic conditions were originally established 
at tree 
level \cite{DiLuzio:2017ogq}. However, since they rely on a mild $\mathcal{O}(10\%)$ tuning, it was 
essential to ascertain 
that renormalization group 
(RG) effects do not undermine the whole construction. 
This was verified 
in Ref.~\cite{DiLuzio:2022tyc},  
where it was shown that RG running 
simply shifts the parameter 
space region where 
the nucleophobic conditions are realised.

In this work, we address another issue concerning the stability of the nucleophobic conditions against a different class of corrections.
The SN 1987A bound  arises from constraints on axion emission from 
a highly dense stellar environment, where  
axion interactions with nucleons are significantly affected by in-medium corrections, as recently demonstrated in Ref.~\cite{Balkin:2020dsr} (see also \cite{Stelzl:2022tcl,Springmann:2023wqn}).
In particular, since the baryonic density $n/n_0$ 
(normalized to the nuclear saturation density $n_0 \simeq 0.16 \, \text{fm}^{-3}$) varies significantly along the 
SN radius (cf.~Fig.~\ref{fig:SN_T-n}), 
even if nucleophobia can be enforced 
locally
for a 
given $n/n_0$, it is unclear whether the  
axion emissivity integrated over the SN core 
would still exhibit a suppression comparable to the one 
observed with axion couplings in vacuum. The purpose of this work is to quantitatively assess to which extent nucleophobic axion models can be regarded as realistic possibilities that endure finite density effects in SNe.

The outline of this paper is as follows. 
In Sec.~\ref{sec:nucleophobinvac}, we review 
nucleophobic axions in vacuum.  
Sec.~\ref{sec:SNFD} provides 
an overview of finite density effects in SNe,  
followed in Sec.~\ref{sec:nucleophobFD} by an analysis of  
finite density corrections to axion-nucleon couplings 
and by an assessment of the fate of nucleophobic axion models
under realistic finite density conditions in Sec.~\ref{sec:revisiting}.
 Finally, in Sec.~\ref{sec:concl} we present our conclusions.

\section{Nucleophobic axions in vacuum}
\label{sec:nucleophobinvac}

Before discussing finite density effects, we begin by introducing axion-nucleon couplings in vacuum, we review the structure of nucleophobic axion models and outline the nucleophobic conditions. 

\subsection{Axion-nucleon couplings in vacuum}
\label{sec:axioncouplinvac}

The axion couplings to nucleons, $N = p,n$, defined via the Lagrangian term 
\beq 
\label{eq:defcN}
c_N \frac{\partial_\mu a}{2 f_a} \bar N \gamma^\mu \gamma_5 N \, , 
\eeq
can be computed in the framework of Heavy Baryon Chiral Perturbation Theory (HBChPT), 
a non-relativistic effective field theory where nucleons are at rest and the axion 
is treated as an external current 
(see \cite{GrillidiCortona:2015jxo,Vonk:2020zfh,Vonk:2021sit} for details). 
In particular, working at leading order with three active flavours, one obtains 
\begin{align}
\label{eq:CpDelta}
c_p &=
\(c_u - \frac{1}{1+z+w} \) \Delta_u 
+ \( c_d - \frac{z}{1+z+w} \) \Delta_d \nonumber \\ 
& + \( c_s - \frac{w}{1+z+w} \)  \Delta_s \, ,
\\
\label{eq:CnDelta}
c_n &=
\(c_u - \frac{1}{1+z+w} \) \Delta_d 
+ \( c_d - \frac{z}{1+z+w} \) \Delta_u \nonumber \\ 
& + \( c_s - \frac{w}{1+z+w} \)  \Delta_s \, ,
\end{align}
where $c_{u,d,s} \equiv c_{u,d,s} (2\GeV)$ 
are low-energy axion couplings to quarks, 
defined analogously to~\eq{eq:defcN}, and 
evaluated 
at the scale $\mu = 2\GeV$ 
by numerically solving the RG equations~\cite{Choi:2017gpf,Chala:2020wvs,Bauer:2020jbp, DiLuzio:2023tqe} 
from the boundary values $c_{u,d,s}(f_a)$
(see below). 
For the quark mass ratios we have 
$z = m_u(2\GeV)/m_d(2\GeV) = 0.465(24)$~\cite{Giusti:2017dmp,MILC:2018ddw}, 
and $w = m_u(2 \GeV)/m_s(2 \GeV) 
=0.0233(9)$ \cite{ExtendedTwistedMass:2021gbo,Bazavov:2017lyh,FermilabLattice:2014tsy,EuropeanTwistedMass:2014osg}. $\Delta_q$, with $q=u,d,s$, are hadronic matrix
elements encoding the contribution of a quark $q$ to the spin operator $S^\mu$ of the proton, 
defined via $S^\mu \Delta q = \langle p | \bar{q} \gamma^\mu \gamma^5 q | p \rangle$.
In particular,
$g_A = \Delta_u - \Delta_d = 1.2754(13)$ 
is extracted 
from $\beta$-decays \cite{ParticleDataGroup:2024cfk},  
while 
$\Delta_u =  0.847(18)(32)$, $\Delta_d =-0.407(16)(18)$ and $\Delta_s = -0.035(6)(7)$ 
(at 2 GeV in the $\overline{\text{MS}}$ scheme) 
are the $N_f = 2+1$ FLAG 2024 averages~\cite{FLAG2023,FlavourLatticeAveragingGroupFLAG:2021npn},
dominated by the results in Ref.~\cite{Liang:2018pis}. 
For further reference, we also define the iso-singlet combination 
$g_0^{ud} = \Delta_u + \Delta_d = 0.44 (4)$. 

Running effects 
on the low-energy  axion couplings  
to light quarks 
can be parametrized as \cite{Choi:2021kuy}
\begin{align}
\label{eq:CuCdCe}
c_{u,d,s} &\simeq c^0_{u,d,s} + r^t_{u,d,s} \, c^0_t \, , 
\end{align}
where $c^0_{u,d,s,t} \equiv c_{u,d,s,t} (f_a)$ are axion couplings defined at the UV scale $\mu = f_a$, 
the parameters $r^t_{u,d,s}$ encode the RG 
correction approximated by taking the leading one-loop top-Yukawa contribution,
and depend logarithmically on the mass scale of the heavy scalar degrees of freedom in the  UV completion of the axion model, 
which is assumed to be of $\mathcal{O}(f_a)$. 
The values of $r^t_{u,d,s}$, 
obtained
by interpolating the numerical solution to the RG equations, 
are tabulated in Appendix B of Ref.~\cite{DiLuzio:2023tqe}. 
In the following, 
we will set as a reference value, 
 $f_a = 10^8$ GeV, 
corresponding to $r^t_{u} = -0.2276$ and $r^t_{d,s} = 0.2290$, 
and we neglect the small logarithmic dependence from 
shifts in $f_a$ 
when varying the axion mass. 

\subsection{Nucleophobic conditions in vacuum}
\label{sec:nucleophobCondinvac}

In order to discuss nucleophobic axion models, in which the 
axion couplings to protons and neutrons are simultaneously 
suppressed, it is convenient to consider the combinations 
\begin{align}
\label{eq:cppcn}
c_p + c_n &= \( c_u + c_d - \frac{1+z}{1+z+w} \) g_0^{ud}  \\
&+ 2 \( c_s - \frac{w}{1+z+w} \) \Delta_s  \nonumber \\ 
&\simeq \( c^0_u + c^0_d -1 \) g_0^{ud} + 2 (c^0_s + r^t_{s} \, c^0_t ) \Delta_s \, , \nonumber \\
\label{eq:cpmcn}
c_p - c_n &=  \( c_u - c_d - \frac{1-z}{1+z+w} \) g_A  \\
&\simeq  \( c^0_u - c^0_d + (r^t_{u} - r^t_{d}) c^0_t - \frac{1-z}{1+z} \) g_A \, , \nonumber
\end{align}
where 
in the last step we neglected $\mathcal{O}(w)$ corrections and 
we have employed the approximation $r^t_{u} + r^t_{d} \simeq 0$ 
(both approximations hold at the per mil level). 
Neglecting also $2 c_s \Delta_s$ in \eq{eq:cppcn}, that amounts 
to  
a few percent (the precise value depending on the specific axion model), 
we see that nucleophobia 
requires the following condition between UV axion couplings 
\beq 
\label{eq:firstnucl}
c^0_u + c^0_d = 1 \, .
\eeq
This condition can be enforced exactly in terms of non-universal PQ charges, and implies $c_p+c_n\simeq 0$.  
In contrast, to ensure that $c_p-c_n\simeq 0$ as well, 
requires tuning $c^0_u - c^0_d$ against the remainder terms in \eq{eq:cpmcn}: 
\beq 
\label{eq:secondnucl}
c^0_u - c^0_d = \frac{1-z}{1+z} - (r^t_{u} - r^t_{d}) c^0_t \, ,   
\eeq
which provides the second nucleophobic condition. 
Note that the residual contribution to the axion-nucleon couplings is eventually set by the $2 c_s \Delta_s$ term in the last line of \eq{eq:cppcn}.

\subsection{Nucleophobic axion models}
\label{sec:nucleophobicmodels}

Different realization of nucleophobic axion models were proposed in Ref.~\cite{DiLuzio:2017ogq} 
(see also \cite{Bjorkeroth:2018ipq,Bjorkeroth:2019jtx,Badziak:2021apn,DiLuzio:2021ysg,Lucente:2022vuo,Badziak:2023fsc,Takahashi:2023vhv,Badziak:2024szg} for model-building variants). 
Here, we consider for definiteness the model denoted as M1 in Ref.~\cite{DiLuzio:2017ogq}, 
although our conclusions apply more generally also to other nucleophobic axion models. 

The M1 model features the same scalar sector of the standard DFSZ model \cite{Zhitnitsky:1980tq,Dine:1981rt},  
namely a complex scalar, singlet under $SU(3)_c\times SU(2)_L\times U(1)_Y$, $\Phi \sim (1,1,0)$ and two Higgs doublets $H_{1,2} \sim (1,2,-1/2)$, which are coupled in 
the scalar potential via the non-hermitian operator $H_2^\dag H_1 \Phi$. The vacuum angle is defined by
$\tan\beta = \vev{H_2}/\vev{H_1}$, so that the 
requirement that 
the PQ current is orthogonal to the hypercharge current
fixes the  PQ charges 
of the two Higgs doublets 
as $\X_1=-\sin^2_\beta \equiv -s^2_\beta$ and
$\X_2= \cos^2_\beta \equiv c^2_\beta$, 
while $\X_\Phi =1$ by normalization. 
The M1 model is further characterized by a 2+1 structure of the PQ charge assignments, namely two generations replicate the same set of PQ charges. Note that, as explained in Ref.~\cite{DiLuzio:2017ogq}, in this case all the entries in the up- and down-type quark Yukawa matrices are allowed and there are no texture zeros. 
In particular, the Yukawa sector of the M1 model contains the following operators
\begin{align}
&\bar q_\alpha  u_\beta H_1\, ,\ \:\bar q_3  u_3 H_{2}\, , \ \:\bar q_\alpha  u_3 H_{1}\, , \ \:\bar q_3  u_\beta H_{2}\, , \nonumber  \\
  \label{eq:m1}
&\bar q_\alpha d_\beta \tilde H_2\, , \ \:\bar q_3  d_3 \tilde H_{1}\, ,\ \:\bar q_\alpha d_3 \tilde H_{2}\, ,\ \:\bar q_3  d_\beta \tilde H_{1}\, , \nonumber \\
&\bar \ell_\alpha  e_\beta \tilde H_1\, , \ \:\bar \ell_3  e_3 \tilde H_{2}\, , \ \, \:\bar \ell_\alpha  e_3 \tilde H_{1}\, ,\ \, \:\bar \ell_3  e_\beta \tilde H_{2}\, , 
\end{align}
where $\alpha,\beta=1,2$ span over first and second generations, 
 while $q,\, \ell$ denote left-handed (LH)  
doublets and $u,\, d,\, e$ right-handed (RH) 
singlets. 
The PQ charges 
stemming from the Yukawa sector 
in \eq{eq:m1} read
\begin{align}
\X_{q_i} &= (0, 0, 1)\, , \nonumber \\
\X_{u_i} &= (s_\beta^2, s_\beta^2, s_\beta^2)\, , \nonumber \\ 
\X_{d_i} &= (c_\beta^2, c_\beta^2,c_\beta^2)\, , \nonumber \\
\X_{\ell_i} &=  -\X_{q_i}\, ,\:\X_{e_i}=-\X_{u_i}\, .
\end{align}
The associated anomaly coefficients 
are $E/N = 2/3$ 
and $2N=1$ (cf.~\cite{DiLuzio:2020wdo} for standard notation), while  
the mixing-independent part of  
the axion couplings to SM fermions are  
\begin{align}
\label{eq:M1coupl}
&c^0_{u,c} =s_\beta^2\, ,\quad 
\; \ \ c^0_t=-c_\beta^2\, , \nonumber \\ 
&c^0_{d,s}=c_\beta^2\, ,\quad 
\; \ \ c^0_b=-s_\beta^2\, , \nonumber \\
&c^0_{e,\mu}=-s_\beta^2\, , \quad 
c^0_\tau=c_\beta^2\, ,
\end{align}
with 
$\tan\beta \in [0.25, 170]$ set by perturbativity \cite{Bjorkeroth:2018ipq}.
Since  the charges of the RH fields are generation independent, there 
are no corrections from RH mixings. In the LH sector mixing effects can play a role because the third generation 
has  different charges from the first two. For the quarks, we assume that the LH mixing matrix 
has CKM-like entries, so that mixing effects are small and will be neglected. 

From \eq{eq:M1coupl} it follows that the first nuclephobic condition in \eq{eq:firstnucl} 
is automatically satisfied, while $c^0_u - c^0_d = s^2_\beta - c^2_\beta$. 
Neglecting RG effects, the second nucleophobic condition in \eq{eq:secondnucl} 
is approximately satisfied for $\tan\beta \simeq \sqrt{2}$. On the other hand, 
as argued in Ref.~\cite{DiLuzio:2022tyc}, RG effects are relevant for the second nucleophobic condition 
and their role is that of changing the cancellation point,  
that in the M1 model gets shifted to $\tan\beta \simeq 1.19$. 
Hence, although the shift in the parameter space region where 
the nucleophobic condition is realised  is sizeable,  
running effects do not prevent the possibility of having nucleophobic models.

The calculation of the axion-nucleon coupling has been done so far assuming zero density. 
However, since the primary relevance of considering nucleophobic axion models lies in their application to the astrophysical environment of SNe, 
it is mandatory to ask 
whether the nucleophobic conditions can still be realised once 
 finite density effects are taken into account.
In the following, 
in order to assess the fate of nucleophobic axion models at finite density,  
we first provide an overview of  density effects in SNe, and then we discuss finite density corrections to axion-nucleon couplings.

\section{Overview of density effects in Supernovae} 
\label{sec:SNFD}

\begin{figure}
    \centering
    \includegraphics[width=1\linewidth]{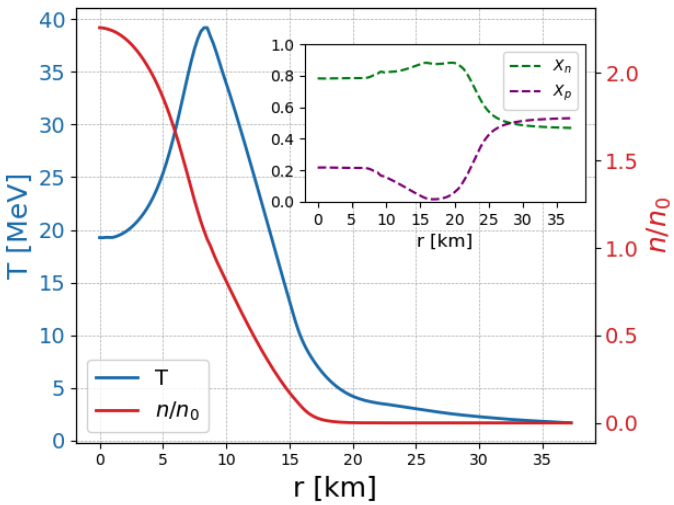}
    \caption{
    Temperature and density profiles of the SN model 
    from Ref.~\cite{SNarchive}, used in this work,  
    as a function of the distance from the center of the star $r$, 1\,s after bounce.
    The inset shows the neutron $(X_n$) and proton $(X_p$) fractions, again as a function of $r$ in km.
    See text for more details.}
    \label{fig:SN_T-n}
\end{figure}

SN cores are extremely dense object, whose baryonic number density, $n=n_p+n_n$, is of the order of the nuclear saturation density $n_0 \simeq 0.16 \, \text{fm}^{-3}$, i.e.~the baryon number density found in 
nuclei. 
The density and temperature profiles of a realistic SN model, 1\,s after bounce, are shown in Fig.~\ref{fig:SN_T-n}, 
including the neutron ($X_n = n_n / n$) and 
proton ($X_p = n_p / n$) fractions.
The model refers to the GARCHING group’s SN model SFHo-s18.8, provided in Ref.~\cite{SNarchive}, based on the {\tt PROMETHEUS-VERTEX} \cite{Rampp:2002bq} code, with the SFHo Equation of State (EoS) \cite{Hempel:2009mc,Steiner:2012rk}. 
The model considers a $18.8M_\odot$ stellar progenitor \cite{Sukhbold:2017cnt} and predicts a neutron star with baryonic mass $1.35M_\odot$.

The production of axions inside a SN can be influenced in 
a non-negligible way by the dense nuclear medium.
Hence, in order to obtain a reliable bound from the cooling argument, it is necessary to determine how finite density corrections affect axion-nucleon interactions, that determine the 
rate of axion emission form SNe.

Certain finite density effects that are quite important in the calculation of the axion emission rate have already been 
identified in previous studies (see, e.g.,~\cite{Carenza:2019pxu,Carenza:2020cis,Lella:2023bfb}). 
For example, if nucleons are sufficiently close to each other, as is the case for densities near $n_0$, the repulsive nuclear forces cannot be neglected and these induce changes in the nucleons dispersion relations
\begin{align}
E_N \simeq m_N+\frac{\left|p_N\right|^2}{2 m_N^*}+U_N \, ,
\end{align}
where $U_N=\Sigma_N^S+\Sigma_N^V$ is the non-relativistic mean-field potential containing contributions of the scalar $\left(\Sigma_N^S\right)$ and vector $\left(\Sigma_N^V\right)$ self-energies (see e.g.~\cite{Hempel:2014ssa}).
The scalar contribution, $\Sigma_N^S$, provides the well-known correction to the nucleon mass
$m_N^* = m_N  + \Sigma_N^S$. 
These effects can be quite significant in the core of a SN, and, for example, can reduce the neutron mass  by about a factor of two. 
However, these calculations 
have not taken into account the modifications to the axion-nucleon couplings that are introduced when the interactions occur inside a highly dense medium.

\section{Nucleophobic axions at finite density}
\label{sec:nucleophobFD}

We proceed now to assess the impact of finite denisty corrections on nucleophobic axion models. First, we review 
the main formalism developed in Ref.~\cite{Balkin:2020dsr} 
for in-medium corrections to axion-nucleon couplings and then 
discuss their impact on axion nucleophobia.

\subsection{Axion-nucleon couplings at finite density}
\label{sec:axioncouplFD}

Finite density corrections to axion-nucleon couplings 
were recently computed in Ref.~\cite{Balkin:2020dsr}. 
There are basically two independent effects to be taken into account: 
the modification of the axion potential due to the change of the chiral condensates 
and the in-medium corrections to the axial couplings $g_A$ (iso-triplet) and $g^0_{ud}$ (iso-singlet). 

The change in the chiral condensates at finite density can be parametrized using the Hellmann-Feynman theorem \cite{Cohen:1991nk}
\begin{equation}
\zeta_{\bar{q}q}(n) \equiv \frac{\vev{\bar{q}q}_n}{\vev{\bar{q}q}_0} = 1 + \frac{1}{\vev{\bar{q}q}_0} \frac{\partial\Delta E(n)}{\partial m_q} \, , 
\label{eq:e30}
\end{equation}
with $q=u, d, s$, 
and the subscripts $n$ and $0$ denoting respectively the in-medium and in-vacuum values. 
Here, $\Delta E(n)$ represents the shift in the QCD ground state energy due to the nucleonic background. 
Neglecting both nuclear interactions and relativistic corrections, one has the 
linear approximation for the chiral condensates,  
$\Delta E = \sum_{N=n, p} m_N n_N$. 
This allows to cast the shift in the condensates in terms of the partial 
derivatives
$\partial m_N / \partial m_q$, 
which are directly related to the so-called sigma terms. 
In particular, one obtains \cite{Balkin:2020dsr} 
\begin{align}
\zeta_{\bar{u} u} (n) &= 1 - b_1 \frac{n}{n_0} + b_2 \[2 \frac{n_p}{n} - 1\] \frac{n}{n_0}\,, \label{eq:e35a} \\
\zeta_{\bar{d} d} (n) &= 1 - b_1 \frac{n}{n_0} - b_2 \[2 \frac{n_p}{n} - 1\] \frac{n}{n_0}\,, \label{eq:e35b} \\
\zeta_{\bar{s} s} (n) &= 1 - b_3 \frac{n}{n_0} \, , \label{eq:e35c} 
\end{align}
where $n = n_p + n_n$ and we defined the $b$-coefficients 
\begin{align}
b_1 &\equiv \frac{\sigma_{\pi N} n_0}{m_\pi^2 f_\pi^2} = 3.5 \times 10^{-1} \(\frac{\sigma_{\pi N}}{45 \, \text{MeV}}\)\,, \label{eq:e36a} \\ 
b_2 &\equiv \frac{\tilde{\sigma}_{\pi N} n_0}{m_\pi^2 f_\pi^2} \frac{m_u + m_d}{m_u - m_d} = -2.2 \times 10^{-2} \(\frac{\tilde{\sigma}_{\pi N}}{1 \, \text{MeV}}\) \, , 
\nonumber \\
b_3 &\equiv \frac{\tilde{\sigma}_s n_0}{m_\pi^2 f_\pi^2} \frac{m_u + m_d}{m_s} = 1.7 \times 10^{-2} \(\frac{\sigma_s}{30 \, \text{MeV}}\) \, ,
\nonumber
\end{align}
and the sigma terms (defined as in Ref.~\cite{Balkin:2020dsr}) have been normalized to their central values \cite{Balkin:2020dsr,Gubler:2018ctz}.
Note that the $\vev{\bar{s}s}_n$ condensate is weakly affected by the nucleonic background,  
while 
$\vev{\bar{u}u}_n \simeq \vev{\bar{d}d}_n$ up to a small isospin correction \cite{Meissner:2001gz}. 

In the following, we will consider only the regime $n < n_c \equiv n_0/b_1 \simeq 2.8 \, n_0$, 
with $n_c$ being the critical density in which one naively expects chiral symmetry restoration in the linear approximation. 
The validity of the linear approximation for the in-medium shift of the chiral condensates 
was estimated to be $n / n_0 \lesssim 1 \div 2$ by including relativistic corrections to the energy of the nucleons \cite{Balkin:2020dsr} 
(see also \cite{Kaiser:2007nv,Goda:2013bka}). 

The change in the chiral condensates at finite density 
can be tracked by 
correcting the quark masses as $m_q \to (\vev{\bar{q}q}_n/\vev{\bar{q}q}_0) m_q$. 
For the axion-nucleon couplings at finite density, 
this implies that one should replace in \eqs{eq:CpDelta}{eq:CnDelta}  $z \to z Z$ and $w \to w W$,
with
\begin{align}
Z &\equiv \frac{\vev{\bar{u} u}_n}{\vev{\bar{d} d}_n} = 1- 2b_2 
\frac{n_n - n_p}{n_0} \, , \label{eq:e49a} \\
W &\equiv \frac{\vev{\bar{u} u}_n}{\vev{\bar{s} s}_n} = 1 - \[b_1 + b_2 \(1 - \frac{2n_p}{n} \) - b_3 \] \frac{n}{n_0} \, , \label{eq:e49b}
\end{align}
where we used $\vev{\bar u u}_0 = \vev{\bar d d}_0 = \vev{\bar s s}_0$ and \eqs{eq:e35a}{eq:e35c}. 
Note that the leading $b_1$ correction has disappeared in the ratio of the condensates in \eq{eq:e49a}. 
Also, in the symmetric-matter limit $n_n=n_p$ (implying $Z = 1$) and neglecting $\mathcal{O}(w)$ terms,  corrections from changes in the chiral condensates vanish. This is at the root of the fact that such effects remain relatively small.

The other effect to be taken into account for the axion-nucleon couplings consists 
in the in-medium correction to the hadronic matrix elements $g_A$ and $g^0_{ud}$, 
while finite-density corrections to $\Delta s$ (whose contribution to $c_{p,n}$ is already subleading)
can be safely neglected. 
In-medium corrections to $g_A$ have been computed in the framework of HBChPT, 
by taking the matrix elements of the space part of the two-body axial-vector currents 
and working in the so-called independent-particle approximation for the background nucleons \cite{Park:1997vv}, 
obtaining (see also \cite{Gysbers:2019uyb,Balkin:2020dsr}) 
\begin{align}
\label{eq:e76}
\frac{(g_A)_n}{g_A} &= 1 + \frac{n}{f_\pi^2 \Lambda_\chi} \\
&\times \left[ \frac{c_D}{4g_A} - \frac{I(m_\pi/k_F)}{3} \left(2 \hat{c}_4 - \hat{c}_3 + \frac{\Lambda_\chi}{2m_N}\right) \right] \, , 
\nonumber
\end{align}
where $(g_A)_n$ denotes the hadronic matrix element at finite 
density,  
$\Lambda_\chi \simeq 700$ MeV is the cut-off scale of the chiral Lagrangian, 
$k_F = (3 \pi^2 n/2)^{1/3} \simeq (270 \, \text{MeV}) (n / n_0)^{1/3}$ is the Fermi momentum,  
\begin{equation}
I(x) = 1 - 3x^2 + 3x^3 \arctan\(\frac{1}{x}\) \, , 
\label{eq:e75}
\end{equation}
and the low-energy constants (LECs) of the HBChPT Lagrangian 
are taken to be \cite{Gysbers:2019uyb} 
\begin{equation}
c_D = -0.85 \pm 2.15 \, , \quad \(2\hat{c}_4 - \hat{c}_3\) = 9.1 \pm 1.4 \, .
\label{eq:e77}
\end{equation}
Using these values, in Ref.~\cite{Balkin:2020dsr} it was estimated
\begin{equation}
\frac{(g_A)_n}{g_A} \simeq 1 - \( 0.3\pm 0.2 \) 
\frac{n}{n_0} \, , 
\label{eq:e78}
\end{equation}
which, however, is valid only for $n/n_0 \ll 1$. 
Hence, in our numerical analysis we will stick to 
the more general expression in \eq{eq:e76}. 

\begin{figure}[t!]
\centering
\quad
\includegraphics[width=0.48\textwidth]{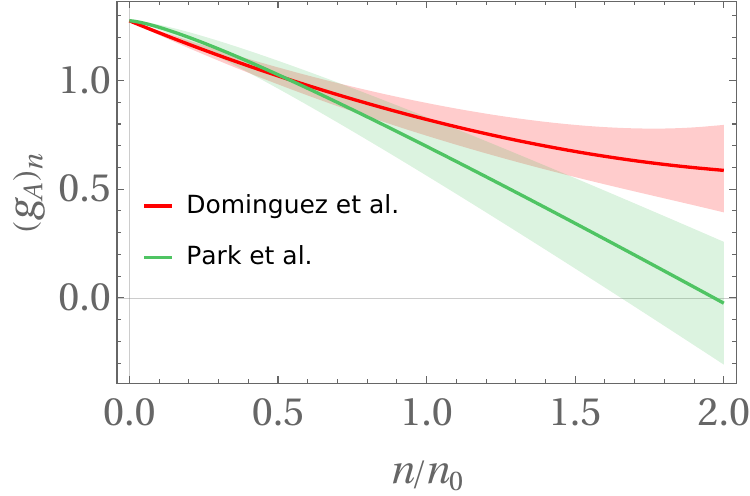}
\caption{Comparison between two different 
determinations of the finite density corrections to $g_A$, 
respectively from Park {\it et al.}~\cite{Park:1997vv} 
and Dominguez {\it et al.}~\cite{Dominguez:2023bjb}.}
\label{fig:comparisongA}
\end{figure}

An alternative derivation of the finite density corrections to $g_A$ was obtained in Ref.~\cite{Dominguez:2023bjb}, 
based on QCD finite energy sum rules. 
This result suggests a weaker dependence of 
$(g_A)_n$ 
on the density. 
The comparison between the abovementioned determinations 
of $(g_A)_n$ is displayed in Fig.~\ref{fig:comparisongA}.
In the following, we will use the 
discrepancy between these two results 
as a further estimate of the theoretical uncertainty.

In principle, one could follow a similar procedure 
as in \eq{eq:e76}
to compute finite density corrections to 
the iso-singlet matrix element $g^0_{ud}$. However, due to the lack of knowledge of the 
associated LECs, we parametrize the finite density correction as
\beq 
\frac{(g^0_{ud})_n}{g^0_{ud}} \simeq 1 + \kappa 
\frac{n}{n_0} \, , 
\label{eq:deltag0ud}
\eeq
with $\kappa \in [-0.3,0.3]$ \cite{Balkin:2020dsr}, in analogy to the finite density correction to $g_A$. 

A final comment on meson condensation is in order. 
At sufficiently high densities, it is expected that a meson condensation phase may occur, leading to significant alterations in hadronic properties (see e.g.~Ref.~\cite{Mannarelli:2019hgn}). 
On the other hand, for symmetric nuclear matter, $n_p = n/2$, such effects are expected to arise for 
$n /n_0 \gtrsim 2$ \cite{Balkin:2020dsr},\footnote{For a related discussion of meson condensation in the SN 
environment, see also \cite{Pons:2000iy,Fore:2019wib}.} 
that is beyond the regime of validity of the linear approximation in 
\eqs{eq:e35a}{eq:e35c}. 
For the SN model considered in this paper (cf.~Fig.~\ref{fig:SN_T-n}), 
at $r\sim 8.5\,$km,  where  the core temperature peaks 
 and axion emission is maximized, $n/n_0 \sim 1$. 
Hence, in the following, we will assume that meson condensation does not occur, which is  justified  for SN models with $n / n_0 \lesssim 2$.

\subsection{Nucleophobic conditions at finite density}
\label{sec:nucleophobCondFD}

We can now proceed to assess the impact of finite density corrections on nucleophobic axion models. 
Following the prescriptions described in \sect{sec:axioncouplFD}, the axion-nucleon coupling combinations in \eqs{eq:cppcn}{eq:cpmcn} 
are modified as follows\footnote{In this analytical argument we neglect $\mathcal{O}(w)$ corrections, 
which are however taken into account in our numerical analysis.} 
\begin{align}
\label{eq:cppcncorr}
(c_p)_n + (c_n)_n &\simeq \( c^0_u + c^0_d -1 \) (g_0^{ud})_n \nonumber \\
&+ 2 (c^0_s + r^t_{s} \, c^0_t ) \Delta_s \, , \\
\label{eq:cpmcncorr} 
(c_p)_n - (c_n)_n &\simeq \Big( c^0_u - c^0_d + (r^t_{u} - r^t_{d}) c^0_t \nonumber \\
& - \frac{1-z Z}{1+z Z} \Big) (g_A)_n \, .  
\end{align}
The crucial point to be observed is that the nucleophobic conditions in vacuum (cf.~\eqs{eq:firstnucl}{eq:secondnucl}) 
are to a good approximation not affected by finite density corrections. 
This is due to the multiplicative nature of the 
corrections 
due to $(g_A)_n$ and $(g_0^{ud})_n$ in \eqs{eq:cppcncorr}{eq:cpmcncorr}, 
as well as to the small effect arising from the $Z$ correction that is proportional to the  
iso-spin breaking $b_2$ coefficient 
and to the size of the asymmetry $(n_n-n_p)/n_0 \lesssim 1$  (cf.~\eq{eq:e49a}).

\begin{figure}[t!]
\centering
\quad
\includegraphics[width=0.45\textwidth]{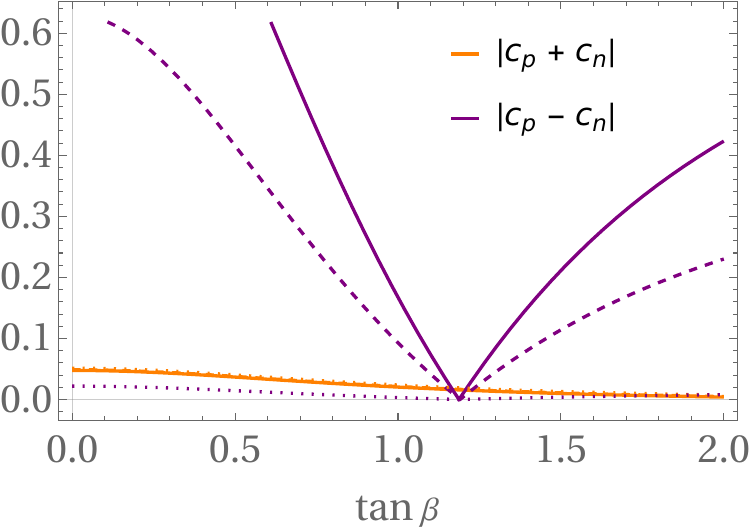} 
\caption{$\abs{c_p + c_n}$ (orange) and $\abs{c_p - c_n}$ (purple, 
with $(g_A)_n$ from Ref.~\cite{Park:1997vv}) 
in the M1 model as a function of $\tan{\beta}$ and for $\kappa=0.3$, for three values of the density, $n/n_0=0$ (solid lines), $n/n_0=1$ (dashed lines), and $n/n_0=2$ (dotted lines). 
\label{fig:cpcntanbeta}
}
\end{figure}

\begin{figure}[t!]
\centering
\quad
\includegraphics[width=0.48\textwidth]{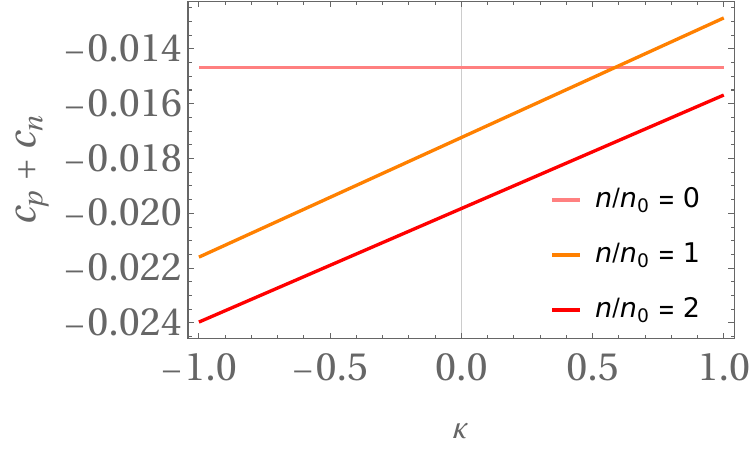}
\caption{{$c_p + c_n$ in the M1 model as a function of $\kappa$, 
for, 
$n/n_0=0$ (pink), $n/n_0=1$ (orange), $n/n_0=2$ (red), and for values of $\tan{\beta}$ corresponding 
for each density to the cancellation  point $c_p - c_n =0$.}  
}
\label{fig:cppcnkappa}
\end{figure}

Our conclusions are confirmed by the 
numerical values of the couplings 
as a function of $\tan\beta$, shown in Fig.~\ref{fig:cpcntanbeta}.  
Remarkably, $i)$ density effects 
lead to a tiny change 
in the cancellation point for $\abs{c_p - c_n}$ 
(not appreciable in Fig.~\ref{fig:cpcntanbeta}), 
in agreement with our analytical argument; 
$ii)$ the lack of a precise knowledge of the factor $\kappa$ in the expression for $(g^0_{ud})_n$ 
(cf.~\eq{eq:deltag0ud}) does not affect the nucleophobic 
solution at the level of the approximations employed in \eq{eq:cppcncorr}. 
Yet, a mild $\kappa$ dependence is reintroduced 
via $\mathcal{O}(w)$ effects, when keeping the full formula in 
\eq{eq:cppcn}. This can be appreciated from Fig.~\ref{fig:cppcnkappa}, where we have plotted 
$c_p+c_n$ for different values of $n/n_0$ and for  
$\kappa$ varying in the wide range  $[-1,1]$. 
Even taking the 
worse case scenario of $\kappa = -1$, 
we find 
$c_p + c_n \sim -0.022$ for $n/n_0 = 1$, 
which still yields a suppression by 
a factor of 20 compared to the KSVZ benchmark, $c_p + c_n \sim -0.43$. 
On the other hand, for relatively large positive values of $\kappa$ the 
nucleophobic condition can even improve compared to the 
zero density case. 

\begin{figure}[t!]
\centering
\includegraphics[width=0.48\textwidth]{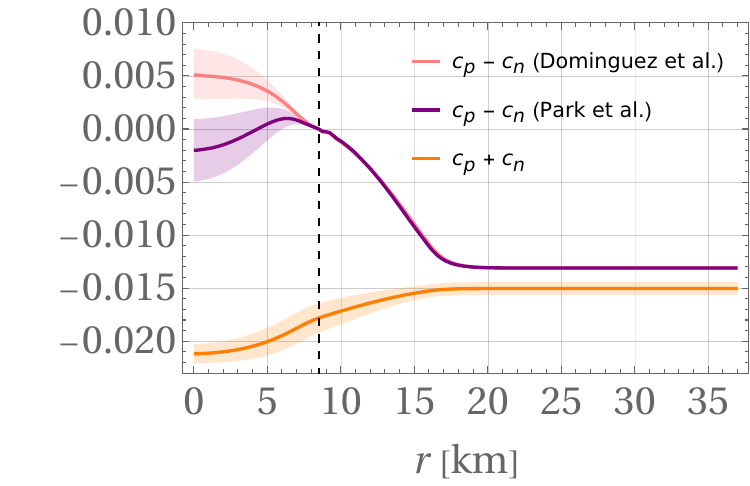}
\caption{Density dependence of $c_p + c_n$ (orange) and $c_p - c_n$ (purple/red) for the M1 model  
with $\kappa=0$, 
and
$\tan\beta$ fixed to the cancellation point of $c_p - c_n$ at $r\simeq 8.5~$km (vertical line). 
The purple line represents the determination of $(g_A)_n$ from Park {\it et al.}~\cite{Park:1997vv}, while the red line corresponds to the determination by Dominguez {\it et al.}~\cite{Dominguez:2023bjb}.  
See text for a description of the parametric uncertainties. 
}
\label{fig:cpcnalongr}
\end{figure}

Finally, Fig.~\ref{fig:cpcnalongr} shows the evolution  
of the values of the coupling combinations $c_p + c_n$ and $c_p - c_n$ 
in the M1 axion model as a function of  the star's radius $r$, 

for the SN density profile depicted in Fig.~\ref{fig:SN_T-n}. 
The variation is due to the change in  density and composition
of nuclear matter. 
The parametric uncertainty 
on $c_p-c_n$ stems dominantly from 
the LECs in 
\eq{eq:e77} (for the case of $(g_A)_n$ taken from \cite{Park:1997vv}) and 
the condensates (for the case of $(g_A)_n$ taken from \cite{Dominguez:2023bjb}),
whereas the uncertainty on $c_p+c_n$ 
is obtained by 
including the error on $g^0_{ud}$  
and by varying 
$\kappa \in [-0.3,0.3]$.
The value of $\tan\beta$ has
been fixed to enforce the cancellation  $c_p - c_n = 0$ at 
$r \simeq 8.5$~km, that 
corresponds to the shell with the highest temperature, 
and hence of maximal axion production. 
Note, however, that due to the  dependence of the axion-nucleon coupling
on the density, this condition can only be applied locally within a thin shell of the SN core.

\section{Revising the SN axion emission for nucleophobic axions}
\label{sec:revisiting}

We have shown that the nucleophobia condition can be maintained at high densities 
and represents a robust feature against density corrections.
However, since the SN core does not have a constant density and the axion-nucleon couplings depend on the local environment, nucleophobia cannot be realised uniformly across the entire core. 
Instead, it can only be enforced within a thin shell where the density and the $n-p$ asymmetry remains approximately constant. 
We refer to this scenario as \textit{localized nucleophobia}.
Given these premises, a key question from a phenomenological perspective is whether the localized nucleophobia condition is sufficient to relax significantly the SN axion bound.

To quantify this, we compare the energy loss due to axions in our nucleophobic model — where nucleophobia is applied in the shell with maximal temperature — against the same nucleophobic model with couplings independent of density.
We perform our numerical analysis using the GARCHING group’s SN model discussed in Sec.~\ref{sec:SNFD} and plotted in Fig.~\ref{fig:SN_T-n}.
For this estimate, we made a set of simplifying assumptions.
First, we neglected the partial degeneracy of nucleons, as most axion emission occurs in the highest temperature regions, where nucleons can be treated as non-degenerate.
Second, we focussed on bremsstrahlung production, excluding potential contributions from pion processes~\cite{Carenza:2020cis,Fischer:2021jfm}, as 
  their exact impact on the 
SN axion emission rate is still uncertain and, in 
any case, we do not expect them to significantly affect our results. 
We have also disregarded the corrections to the bremsstrahlung rate that have been recently discussed in the literature (see Ref.~\cite{Carenza:2019pxu}), 
as well as the contribution to the axion emission from strange matter \cite{Cavan-Piton:2024ayu}.

Since our primary objective is to compare the axion luminosity in two distinct cases using the same set of assumptions, these approximations are not expected to  affect significantly our results,  
and we are confident that our main conclusions remain robust.
Lastly, we have performed an integration by summing up the contributions of $c_{n}$ over neutron density and of $c_{p}$ over proton density, provided by our numerical models (cf. Fig.~\ref{fig:SN_T-n}). 
We ignored interference terms of the form $c_{n}c_{p}$, which add unnecessary complexity and are somewhat suppressed in their contribution to the emission rate~\cite{Lella:2022uwi}.

Using these approximations, the energy emitted in axions per unit time can be calculated through the volume integral~\cite{Raffelt:1996wa,Giannotti:2005tn,Raffelt:1993ix}:
\begin{equation}
\label{eq:luminosity}
L_a(t) \propto \int_0^R \bar c_{N}^2 (r) \rho^2(r) T^{7/2}(r) \,r^2 dr \, ,
\end{equation}
where $\bar c^2_{N} = \left(c_{n}X_n\right) ^2 + \left(c_{p}X_p\right) ^2$
is an effective axion-nucleon coupling 
which takes into account the different densities of protons and neutrons, and $R$ is the star's radius. 
Notice that all quantities under integration in the previous equation depend on time, and that 
constant factors have been omitted as they cancel out when comparing luminosities between axion models. 

\begin{figure}[t!]
    \centering
    \includegraphics[width=1\linewidth]{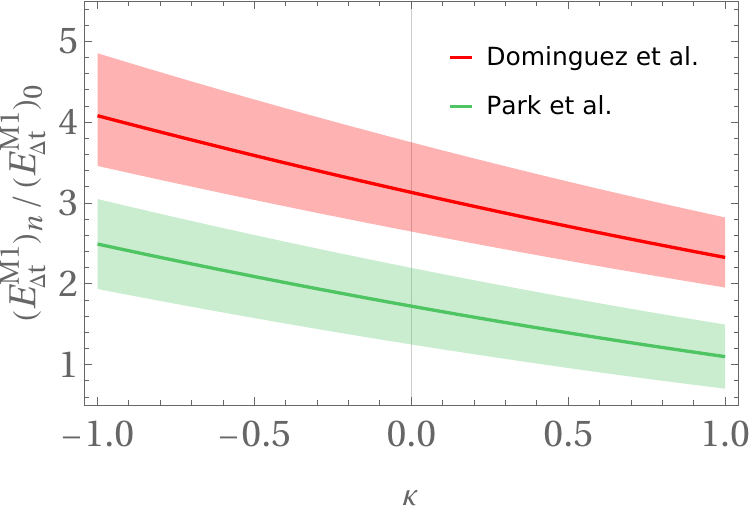}
    \caption{{SN axion  luminosity integrated between $t=1\,$s and $4$\,s  for the nucleophobic M1 model 
    as a function of $\kappa$, 
    normalized to the zero density result.
    The value of $\tan\beta$ is chosen to optimize     
    nucleophobia  respectively in the shell with maximum temperature at $t=1$ s, and in vacuum. 
    The two lines correspond to 
    the 
    calculation of $(g_A)_n$ 
    in Park {\it et al.}~\cite{Park:1997vv} 
    and Dominguez {\it et al.}~\cite{Dominguez:2023bjb}}.}
    \label{fig:luminosityM1}
\end{figure}
In Fig.~\ref{fig:luminosityM1} we plot
the results of the  integration of the luminosity \eq{eq:luminosity} over the  interval $\Delta t =[1-4]$\,s  post bounce, during which the axion emission is maximal.\footnote{For $t\lesssim 1\,$s relatively low values of the core temperature 
suppress the emissivity~(see e.g.~fig.~7 in Ref.~\cite{Ott:2012mr} or  Fig.~11 in Ref.~\cite{Sumiyoshi:2005ri}). For  $t\gtrsim 4\,s$ the temperature quickly drops suppressing again the emissivity.} 
The figure depicts the SN axion integrated luminosity 
$\left(E^{\text{M1}}_{\Delta t}\right)_n$
for the nucleophobic M1 model,  
normalized to the lumionsity  $\left(E^{\text{M1}}_{\Delta t}\right)_0$ for the same model without finite density corrections, 
as a function of the parameter $\kappa$. 
%
%
In evaluating $\left(E_{\Delta t}^{\mathrm{M} 1}\right)_0$,  $\tan\beta$ has been fixed to the value that optimizes nucleophobia in vacuum, while for  
  $\left(E_{\Delta t}^{\mathrm{M} 1}\right)_n$ to the value that 
optimizes nucleophobia in the shell with maximal temperature at $t\simeq 1\,$s.
To quantify the theoretical uncertainty, we display two different cases for $(g_A)_n$, corresponding to the calculation in 
Dominguez {\it et al.}~\cite{Dominguez:2023bjb} (red line) and in 
Park {\it et al.}~\cite{Park:1997vv} (green line),  
where the green band depicts 
the parametric uncertainty from the LECs in \eq{eq:e77}.
The difference between the Park {\it et al.}~and Dominguez {\it et al.}~results, shown in the figure, is a consequence of the different behavior of $g_A$ at high density predicted by the two models, as shown in Fig.~\ref{fig:comparisongA}. 
This becomes particularly important at late times, $t\gtrsim 1$ s, when the region of maximal axion production (higher $T$) moves toward the central regions, where the density is higher.

As is evident from the figure,  finite density effects tend to reduce the overall level of nucleophobia, 
since the luminosity is slightly increased.
This is not surprising, since 
in the realistic case, where only the localized nucleophobic condition 
can be imposed, nucleophobia is less effective than in the vacuum case, 
where it is possible to impose the nucleophobic condition globally.
Nevertheless, the axion time-integrated luminosity calculated with the full density corrections and a realistic SN density profile is only about a factor of 2-4 
larger  compared to the value obtained when finite density effects are neglected (and even less for certain values of $\kappa$).
This corresponds to a bound on the axion mass that is 
a few times stronger compared to the result $m_a\lesssim 0.20\,$eV  obtained in Ref.~\cite{DiLuzio:2017ogq}. 
The dependence of our result on $\kappa$ and, in particular, the fact that the ratio between the luminosity in the two cases is suppressed at large $\kappa$, 
stems from the effects of this parameter on the corrections to the axion couplings. 
In fact, as discussed in Sec.~\ref{sec:nucleophobFD}, large values of $\kappa$ may actually reduce the absolute value of the couplings (cf. Fig.~\ref{fig:cppcnkappa}).

\section{Conclusions}
\label{sec:concl}

Axion nucleophobia was first proposed in Ref.~\cite{DiLuzio:2017ogq} to address the stringent SN 1987A constraints on QCD axions, which significantly restrict the  parameter space relevant to current and future 
axion search experiments (see, e.g., Refs.~\cite{Irastorza:2018dyq,Sikivie:2020zpn,IAXO:2019mpb,DiLuzio:2021ysg,Caputo:2024oqc}). 
These constraints stem from the axion-nucleon couplings. By suppressing this interaction, nucleophobic models~\cite{DiLuzio:2017ogq}
allow to expand significantly the viable parameter space, opening up opportunities for new experimental searches.

In this study, we examined the impact 
of finite density effects
on nucleophobic axion models, focusing on the high-density SN core 
environment. We demonstrated that the nucleophobic condition, which suppresses axion couplings to nucleons, is maintained under in-medium corrections up to supersaturation densities. Additionally, we showed that 
after the inclusion of finite density effects, nucleophobic axion models still enable a significant relaxation of the stringent SN 1987A axion bound. In particular,   with respect to the original (zero density) estimate in Ref.~\cite{DiLuzio:2017ogq}, 
the bound on the axion mass gets  strengthened  by only a  
factor between $2$ and $4$, depending on the specific  
model assumed for $g_A$ 
(cf.~Fig.~\ref{fig:comparisongA}).
Although the relaxation of the axion bound 
is 
less significant compared to the vacuum case
(or to the case in which the density and composition 
of nuclear matter is held constant at certain values) 
this is a notable result since, as discussed in the text, nucleophobia cannot be imposed throughout the entire SN core 
and at all times,  
due to its non-uniform density and composition
as well as to the time variations during  proto-neutron star  core evolution. 
In other words, the softening of the nucleophobic condition due to 
the integration over the non-homogeneous SN environment, and over the time interval most relevant for axion emissivity, only partially hinders the relaxation of the SN axion bound.
Therefore, nucleophobic axions continue to be a viable possibility for loosening  the SN axion bound, and remain compelling targets for future axion searches.

\section*{Acknowledgments}
{We warmly thank  Thomas Janka for giving us access to the {\tt GARCHING} group archive. 
We also thank 
Stefan Stelzl 
for useful communications, 
and 
Cristian Villavicencio 
for clarifications about Ref.~\cite{Dominguez:2023bjb}.
The work of LDL and FM is supported
by the European Union -- Next Generation EU and
by the Italian Ministry of University and Research (MUR) 
via the PRIN 2022 project n.~2022K4B58X -- AxionOrigins.  
FM and EN are  supported in part by the INFN “Iniziativa Specifica” Theoretical Astroparticle Physics (TAsP). 
 The work of EN is also supported  by the Estonian Research Council grant PRG1884. EN also acknowledges partial support from the CoE grant TK202 “Foundations of the Universe” and from the CERN and ESA Science Consortium of Estonia, grants RVTT3 and RVTT7. 
MG acknowledges support from the Spanish Agencia Estatal de Investigación under grant PID2019-108122GB-C31, funded by MCIN/AEI/10.13039/501100011033, and from the “European Union NextGenerationEU/PRTR” (Planes complementarios, Programa de Astrofísica y Física de Altas Energías). He also acknowledges support from grant PGC2022-126078NB-C21, “Aún más allá de los modelos estándar,” funded by MCIN/AEI/10.13039/501100011033 and “ERDF A way of making Europe.” Additionally, MG acknowledges funding from the European Union’s Horizon 2020 research and innovation programme under the European Research Council (ERC) grant agreement ERC-2017-AdG788781 (IAXO+).
This article/publication is based upon work from COST Action COSMIC WISPers CA21106, supported by COST (European Cooperation in Science and Technology).

\bibliographystyle{apsrev4-1.bst}
\bibliography{bibliography}

\end{document}